\documentclass{he_symp}
\usepackage{psfig,graphicx,epsfig}
\usepackage{color}
\usepackage{amsmath,amssymb,epic,eepic,array}
\begin{document}
\renewcommand{\FirstPageOfPaper }{ 112}\renewcommand{\LastPageOfPaper }{ 113}

\title{Vortex Configurations, Oscillations and Pinning in Neutron Star Crusts}
\author{M.Hirasawa\inst{1} \and N.Shibazaki\inst{2}}  
\institute{Department of Physics, Rikkyo University, Nishi-Ikebukuro, Tokyo 171-8501, Japan\and  Department of Physics, Rikkyo University, Nishi-Ikebukuro, Tokyo 171-8501, Japan}
\maketitle

\begin{abstract}
The rotating neutron superfluid in the inner crust of a neutron star is threaded by quantized vortex lines. The pinning force from lattice nuclei and the Magnus force from neutron superfluid act onto a vortex line that has a finite tension. The configuration of a vortex line in equilibrium comprises a number of pinned straight lengths separated by unpinned kinks when the rotation axis is slightly inclined to the major axis of a crystal lattice and the Magnus force is not so strong. Energy of $\sim 4 \mbox{ MeV}$ is required to form a kink at densities of $\sim 3.4 \times 10^{13} \mbox{ g cm}^{-3}$. Magnus force makes the vortex line parabolic on global scale. There exist two modes, rotational and helical (Kelvin), for oscillations allowed on a vortex line. The vortex oscillations in the pinned straight segments are possible only above a minimum frequency. We find no unstable mode that grows with time. Hence, the vortex configurations with kinks may be stable and yield the strongly pinned state. The essence of our results may not be altered even when we consider the vortex behaviors in a polycrystalline structure. Our studies suggest that pinning of a vortex line in a polycrystalline structure would be still strong enough to explain the large
glitches.
\end{abstract}

We show the equilibrium configuration of a vortex line in Fig. 1, where we adopt the coordinates fixed in the lattice nuclei. The z-axis is chosen to be parallel to one of the major axes of a body-centered cubic lattice and the neutron superfluid is assumed to flow in the y-direction. The horizontal coordinate $\phi$ expresses displacement of a vortex segment from the z-axis. All coordinates are multiplied by $g = 2\pi/a$, where $a$ is the lattice constant. Here the Magnus force is ignored for simplicity. The numbers on the curves denote the inclination angle between the rotation and main crystal axes. We find that when the inclination angle is small, vortex lines in equilibrium configurations move from one lattice plane to the next by forming a kink and a number of kinks increases with increasing inclination angle. Note that the average orientation of a vortex line corresponds to the direction of the rotation axis. A vortex line can orient its average direction to the direction of the rotation axis by adjusting a number of kinks. 
\begin{figure}
\centerline{\psfig{file=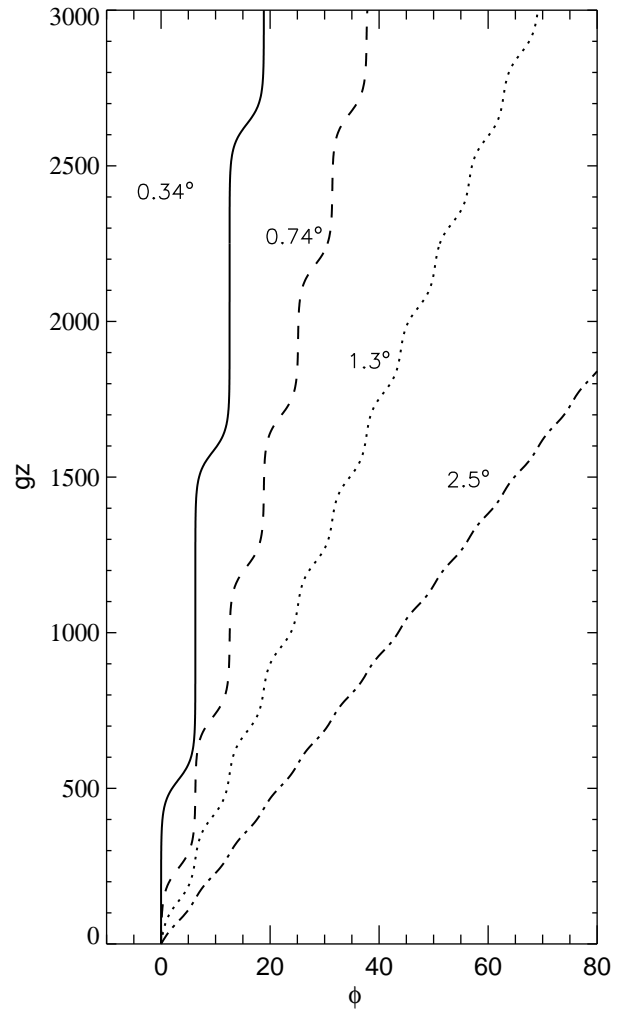,width=8.8cm,clip=} }
\caption{ Vortex equilibrium configurations with kinks for different inclination angles of a rotation axis against the z-axis. Here the Magnus force is ignored. The numbers on the curves denote the inclination angle. The average orientation of a vortex line corresponds to the direction of a rotation axis of the star. The main lattice planes are located at the horizontal coordinate of $\phi=2n\pi$.
\label{fig1}}
\end{figure}
 Apart from the kink parts, the vortex lines are straight and in parallel to the z-axis, and are firmly pinned to the lattice nuclei. Even though the extra energy is required to create kinks, the total energy
of a vortex line with kinks is much lower than that of the inclined straight vortex line without kinks since except for kink parts the vortex line lies in the bottom region of the pinning potential.

\begin{figure}
\centerline{\psfig{file=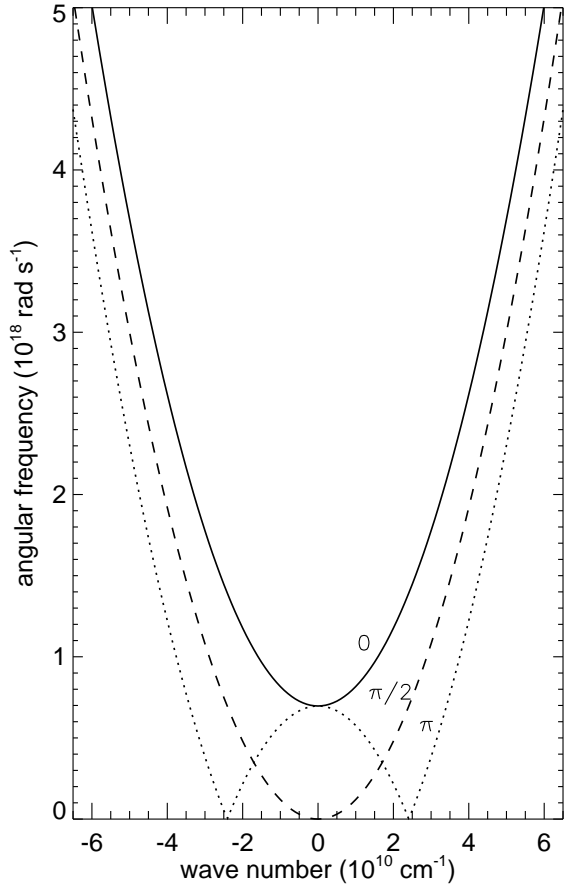,width=8.8cm,clip=} }
\caption{Dispersion relations for oscillations allowed on a vortex line. The numbers by the curves denote displacement of the vortex segment, $\phi$, from the z-axis in equilibrium configuration. The oscillations in the pinned segments are possible only above a minimum frequency.
\label{fig2}}
\end{figure}

Figure 2 illustrates the dispersion relations for oscillations excited on a vortex line.  The numbers by the curves denote a displacement of the vortex segment, $\phi$, in equilibrium. We find that there exist two oscillation modes, rotational and helical (Kelvin) modes. At lower wave numbers the angular frequency of oscillations is determined mainly by the rotational mode, while at higher wave numbers mainly by the Kelvin mode. We should note that there exists the minimum frequency for the oscillations especially when $\phi$ is close to $2n\pi$. The oscillations with frequencies lower than the minimum cannot be excited on and cannot propagate in the pinned part of a vortex line, whereas the oscillations with frequencies higher than the minimum can propagate along a vortex line, varying the wave length.

Recently, Jones (1998) argues that vortex interaction with a polycrystalline structure does not provide pinning strong enough to explain the large glitches observed in the Vela pulsar, considering the kink motion along a vortex line. Our results, however, show that the vortex
equilibrium configurations with static kink structure are stable. Hence, the kink motion as required by Jones is less likely. The kink solution can be expressed as a sum of Fourier components of different wave numbers. The dispersion relation shows that the phase velocity of vortex waves depends on the wave number. Hence, even if a kink is formed and starts to move, the kink feature will be smeared out during propagation. The vortex equilibrium configuration is composed of the static kink and straight segments. A vortex line in equilibrium lies deep in the pinning potential well and is strongly pinned to the lattice nuclei in its most part. We conclude from these results that pinning may be strong enough to explain the large glitches observed in the Vela pulsar.


\clearpage

\end{document}